\documentclass{ws-p8}

\newcommand{\beq}{\begin{equation}}
\newcommand{\eeq}{\end{equation}}
\newcommand{\ba}{\begin{array}}
\newcommand{\ea}{\end{array}} 
\newcommand{\beqa}{\begin{eqnarray}}
\newcommand{\eeqa}{\end{eqnarray}}
\newcommand{\dis}{\displaystyle}
\newcommand{\no}{\nonumber}
\newcommand{\cL}{{\cal L}}

\newcommand{\da}{^\dagger}
\newcommand{\lsim}{\stackrel{<}{_\sim}}
\newcommand{\gsim}{\stackrel{>}{_\sim}}
\newcommand{\eps}{\varepsilon}
\newcommand{\epsp}{\varepsilon'}

\begin{document}

\begin{flushright}
{\small CERN-TH/2000-327}
\vskip 0.8 cm
\end{flushright}

\title{\uppercase{Weak decays of pseudo Goldstone bosons}\footnote{To appear in the 
proceedings of {\em Chiral Dynamics 2000}, Newport News, USA, 17-22 July 2000.}}

\author{Gino Isidori}

\address{Theory Division, CERN, CH-1211 Geneva 23, 
Switzerland\footnote{On leave from INFN, Laboratori Nazionali di 
Frascati, Via Enrico Fermi 40, I-00044 Frascati (Rome), Italy.}
\\E-mail: Gino.Isidori@cern.ch}

\maketitle

\abstracts{A short overview of recent progress 
in describing kaon and pion decays 
within Chiral Perturbation Theory is presented. 
Particular attention is devoted to the issues of 
final-state interactions and isospin breaking 
in $K \to 2 \pi$, as well as to the estimate of 
long-distance contributions in 
$K \to \ell^+\ell^-  (\pi)$.}

\section{Introduction}
The weak decays of light pseudoscalar mesons,
namely the decays of kaons and charged pions, 
represent the widest domain of applicability 
of Chiral Perturbation Theory (CHPT). 
The field is certainly dominated by kaon decays, 
with about 50 different channels observed 
so far, which constitute by themselves one of the richest 
sources of information about the Standard Model 
as a whole.

The interest and the variety of such decays is 
definitely too large to be covered in one talk.
I will therefore concentrate on few selected 
topics, not covered by other plenary speakers, 
among those where interesting work
has recently been done.
In the next section I will review 
the development in the construction of 
chiral Lagrangians without baryon fields.
Two {\it hot topics} in $K \to 2 \pi$ decays,
namely the issues of final-state interactions 
and isospin breaking, will be discussed in 
section 3. Section 4 is devoted to 
the rare processes $K \to  \pi \ell^+\ell^-$
and  $K_L \to \ell^+\ell^-$,
and the last section contains
some concluding remarks. 

\section{Developments in the construction of 
chiral Lagrangians}

Kaon and charged-pion decays are 
usually classified into two large categories: 
(semi-)leptonic (including all pion decays)
and non-leptonic. 
As long as electromagnetic interactions are 
neglected, the two categories have very different 
properties from the point of view 
of chiral dynamics:
\begin{itemize}
\item 
In semileptonic transitions,
where the lepton pair interacts with the 
mesons only via the $W$ boson, the latter   
can be considered as an external field. 
These transitions can therefore be simply 
described in terms of the {\it strong}
generating functional of CHPT, with an appropriate 
identification of the external sources.\cite{GL}
\item 
In non-leptonic decays, where quarks lines are attached 
to both vertices of the $W$ propagator, the 
latter cannot be considered as an external field.
These processes can be described within CHPT
introducing a new generating functional,
that transforms linearly under chiral rotations 
as the four-quark $|\Delta S|=1$ effective 
Hamiltonian.\cite{KMW}
\end{itemize}
The structure of both functionals has been known 
up to $O(p^4)$, i.e.~to the next-to-leading 
order, for several years, allowing a systematic 
inclusion of meson--meson interactions up to one 
loop in both types of processes.\cite{GL,KMW}
The number of low-energy couplings (LECs) appearing at 
$O(p^4)$ is apparently large: 12 in the strong 
sector and 37 in the octet part of the weak
non-leptonic functional. Nonetheless the 
theory still has a good predictive power 
since only few LEC combinations appear 
in physical processes.\cite{HD}
More recently, also the $O(p^6)$ structure of the 
strong generating functional has been 
determined,\cite{BCE} leading to a systematic
inclusion of mesonic two-loop effects 
in semileptonic processes. The total number of  
$O(p^6)$ strong LECs is definitely large (about 
one hundred), but a few accurate predictions can 
still be made concerning specific observables.\cite{Gil} 

The distinction between semileptonic and 
non-leptonic processes becomes less clear when
electromagnetic interactions are taken 
into account. Indeed the $W$ boson 
can no longer be considered as an external source 
if the charged lepton interacts electromagnetically 
with the pseudoscalar fields. On the other hand
electromagnetic corrections to the leading $O(p^2)$ terms 
cannot be neglected with respect  to the 
purely mesonic two-loop effects.
As we shall discuss in section 4, 
electromagnetic interactions play an even more 
important r\^ole in non-leptonic transitions,
where they could compete already with the $O(p^2)$ 
terms in $\Delta I=3/2$ transitions.

The first step toward a systematic inclusion of
electromagnetic effects has been undertaken 
almost five years ago,\cite{Urech}
with the determination 
of the $O(e^2p^2)$ local terms needed to 
regularize the one-loop divergences induced by 
virtual photons in the strong sector. 
This step, however, was not sufficient for a complete 
treatment of $O(e^2p^2)$ effects, be it in
semileptonic or in non-leptonic decays. 
Interestingly, the missing ingredients for 
this program have recently become available:
Knecht {\em et al.}\cite{Knecht1}
have determined the additional divergences (and 
associated counterterms) generated at one loop 
in the strong sector by the presence of 
virtual leptons. Finally the 
effect of virtual photons in non-leptonic
$|\Delta S|=1$ transitions has been taken into 
account in a systematic way up to $O(e^2p^2)$.\cite{Cirigliano1,noi}

Given this situation, we can state that the structure 
of chiral Lagrangians is known, at present, to a 
very good accuracy for almost all decays of 
phenomenological interest. Nonetheless this is only one 
side of the problem, the other being the 
determination of the LECs. As I will discuss in the 
following, substantial improvements are still needed 
in the latter direction in order to match  
the potential degree of precision allowed by the 
present knowledge of chiral Lagrangians.

\section{$K\to 2 \pi$}
Neglecting isospin-breaking (and electromagnetic) 
effects, $K\to 2 \pi$ amplitudes can be decomposed as follows
\beqa
A(K^0 \to \pi^+\pi^-) &=& A_0 e^{i \delta_0} +
                          {1\over \sqrt{2}} A_2 e^{i \delta_2}~, \no \\
A(K^0 \to \pi^0\pi^0) &=& A_0 e^{i \delta_0} -
                          \sqrt{2} A_2 e^{i \delta_2}~, \no \\
A(K^+ \to \pi^+\pi^0) &=& { 3 \over 2 } A_2 e^{i \delta_2}~, 
\label{Adec1}
\eeqa
where $A_I$ are the weak amplitudes (real in the absence of CP 
violation) 
and $\delta_I$ denote the $S$-wave strong phases of the  
$\pi\pi$ system in isospin $I$. As  is well known, 
the ratio $\omega=|A_2/A_0|$ is found experimentally to be 
very small ($\omega\sim 1/22$), and a precise dynamical 
explanation of this result (the $\Delta I=1/2$ rule) is
%still missing.
one of the most challenging open problems in understanding 
non-leptonic weak decays.

The effective $|\Delta S|=1$ four-quark Hamiltonian 
contributing to these processes contains operators 
that transform as $(8_L,1_R)$ and $(27_L,1_R)$ 
under chiral rotations.\footnote{~Consistently with 
the assumption of neglecting isospin-breaking effects,
at this level we can neglect the $(8_L,8_R)$ term
induced by electromagnetic interactions.} 
Both these structures have a unique chiral realization at $O(p^2)$,
which leads to the following simple Lagrangian:\footnote{~Notation 
is as in D'Ambrosio {\em et al.}\protect\cite{HD} }
\beqa
{\cL}_W^{(2)} & = & G_8 F^4 \left[ D_\mu U\da D^\mu U \right]_{23}  \no \\ 
&+& G_{27} F^4 \left[ (U\da \partial_\mu U)_{23}(\partial_\mu U\da U)_{11} 
+{2 \over 3} (U\da \partial_\mu U)_{21}(\partial_\mu U\da U)_{13}  \right]+
{\rm h.c.}\quad
\eeqa
The dimensional couplings $G_8$ and $G_{27}$ are free 
parameters from the point of view of pure CHPT, which by 
itself is not expected (at least at the lowest order) to
shed light on the origin of the $\Delta I=1/2$ rule. 
The latter is phenomenologically implemented 
%fitting the expressions
%\beq
%A_0^{(2)} = \sqrt{2} F^2 
%            \left( G_8 +\frac{1}{9} G_{27} \right) (m_K^2-m_\pi^2) ~, \quad 
%A_2^{(2)} = \frac{10}{9}  F^2  G_{27} (m_K^2-m_\pi^2)
%\eeq
%from data, i.e. employing the values 
by fitting $G_8$ and $G_{27}$ from  $K\to 2 \pi$ amplitudes:
\beq 
\left| G_8 \right|^{\rm exp}_{\pi\pi} = 9.1 \times 10^{-6}~{\rm GeV}^{-2}~, \qquad 
\left| G_{27} / G_8 \right|^{\rm exp}_{\pi\pi} \simeq {1 / 18}~.
\label{G8Kpp}
\eeq
Once $G_8$ and $G_{27}$ have been fixed,
CHPT becomes predictive and efficient in describing 
other processes, such as  $K\to 3 \pi$ decays.

The puzzle of the  $\Delta I=1/2$ rule arises in the attempt to 
derive (\ref{G8Kpp}) from the underlying SM dynamics. Indeed,
in the strict $N_c\to \infty$ limit of QCD, when the dominant 
operators in the effective four-quark Hamiltonian have no anomalous 
dimension and their hadronic matrix elements can be factorized, 
one finds:
\beq 
\left| G_8 \right|^{\rm th}_{N_c \to \infty} = 
\left| G_{27} \right|^{\rm th}_{N_c \to \infty} =  1.1 \times 10^{-6}~{\rm GeV}^{-2}~.
\eeq
The situation slightly improves when QCD corrections to the Wilson 
coefficients of the effective four-quark Hamiltonian 
are taken into account. Evaluating the latter the leading 
order with a renormalization scale $\mu\sim 1$~GeV, 
and still employing factorized 
matrix elements, leads to an enhancement factor $\sim 2$ 
for $G_8$ and a suppression of around $0.7$ for $G_{27}$.
This estimate is not very precise, since
the matrix elements of the four-quark operators 
($\langle Q_i \rangle$)
do not have,   at this level, the correct scale 
dependence needed to match the one of the Wilson 
coefficients; nonetheless it shows that the bulk 
of the effect is still missing. In particular an 
enhancement factor of about $4$ is still needed in 
the $\Delta I=1/2$ amplitude. If the Wilson 
coefficients are evaluated at a scale $\mu \gsim 1$~GeV,
condition necessary to trust their perturbative 
estimate, this enhancement can only be addressed to 
matrix elements 
that differ substantially from their factorized values.

\subsection{Final-state interactions}
The evaluation of hadronic matrix elements 
is a non-perturbative problem related mainly 
to low-energy dynamics. It is therefore natural 
to address the question whether CHPT can help 
to clarify it, once effects beyond $O(p^2)$
are taken into account.

An interesting suggestion in this direction has 
recently been proposed by Pallante and Pich,\cite{PP}
following an older work by Truong.\cite{Truong}
Their proposal is based on the observation that 
the large $\pi\pi$  scattering phase in $I=0$ 
signals a large final-state interaction (FSI) effect, 
not well described by CHPT at lowest order,
where there is no absorptive contribution to the 
decay amplitudes. 
This observation is certainly correct, as it can be checked 
by the explicit $O(p^4)$ CHPT calculation of 
Kambor {\em et al.}.\cite{KMW2} There, it has been 
shown that pion loops provide a sizeable renormalization
of $G_8$. The amount of this 
effect, however, cannot be unambiguously determined 
within CHPT alone, owing to the presence of  $O(p^4)$
local operators with unknown 
couplings.\footnote{~Assuming for instance that the effect 
of the local operators is negligible at a given CHPT 
renormalization scale $\mu_\chi$ (not to be confused with 
the renormalization scale of the four-quark Hamiltonian),
and varying this scale between 0.5 and 1 GeV,
one finds that pion loops provide an 
enhancement of $A^{(4)}$ over $A^{(2)}$
that ranges between 
20\% and 80\%.\protect\cite{IP}}
The further assumptions employed by Pallante 
and Pich in order to obtain a quantitative  
information about the $\langle Q_i \rangle$ 
are that  i) FSI effects can be 
unambiguously re-summed to all orders in CHPT using 
an Omn\`es factor,\cite{Truong}  
ii) in some cases (most notably in the case of the 
operator $Q_6$ relevant to $\epsp/\eps$) 
this FSI effect constitutes 
the dominant correction with respect to the 
large $N_c$ estimate of the matrix element. 

The use of dispersion relations (and the corresponding 
Omn\`es solution) provides, in some cases, an efficient 
tool to resum FSI effects. However this is hardly 
justified in the context of $K\to\pi\pi$ amplitudes.\cite{BCFIMS}
In order to apply this tool to $K\to\pi\pi$ one could try to 
treat $m_K$ and $s=(p_{\pi_1}+p_{\pi_2})^2$ as two independent 
variables, but in this case one would have to deal with 
the (unknown) effect of operators that are eliminated 
by the equations of motion. Otherwise, identifying $s$ and $m^2_K$, 
one should be worried by the impact induced on $\pi\pi$ phases
by a variation of the kaon mass.
 
Besides the justification of this approach, an even more serious 
problem is posed by the fact that the Omn\`es solution does not 
completely determine the amplitudes. In the most optimistic 
case it can be used to relate the physical amplitudes 
to those obtained at a different value of $s$.\cite{BCFIMS} 
Not surprisingly,  this problem is somehow equivalent to the issue 
of determining 
the local $O(p^4)$ terms within CHPT. Thus we are  back to the 
starting point,  unless we make the further assumption that large 
$N_c$ provides a good estimate of some $\langle Q_i \rangle$
at $s=0$. Again it is hard to find a 
justification of this hypothesis, in particular it is not
clear at all whether large $N_c$ should work better at 
$s=0$  than at large $s$. The argument that chiral 
corrections are small at $s=0$ does not help since one 
should rather be worried by the size of $1/N_c$ corrections. Moreover, 
if $s\not=m^2_K$, one could still expect sizeable chiral corrections 
driven by $m_K$ rather than $s$ that are potentially large at 
$s=0$. Finally, we note that applying this procedure 
to all the  $\Delta I=1/2$ operators  {\em does not} lead to 
reproducing the 
observed $|A_0|$. Indeed the FSI enhancement 
factor computed in this way\cite{PP} is only $\sim 1.5$,
to be compared with the factor 4 needed by the data. 
It is thus clear that, at least in some cases, 
this procedure is incomplete.

Having pointed out these problems, it should also be stressed 
that the work of Pallante and Pich\cite{PP} had the merit of 
focusing the attention on a potentially large effect,
sometimes ignored in the literature, which needs to be 
taken into account when trying to evaluate $K\to\pi\pi$
matrix elements of four-quark operators.  I believe that 
a satisfactory solution to this problem can only be
achieved within a self-consistent 
calculation of the full dispersive 
contribution to the $\langle Q_i \rangle$. 
Several attempts in this direction exist
in the literature, using both analytical 
tools and lattice QCD. In the first category 
there have been encouraging results (obtained by means of different 
strategies) concerning 
the $\Delta I=1/2$ rule\cite{DeltaI} and the 
matrix elements  of 
electroweak operators.\cite{deRafael} 
Nonetheless all the analytical methods are still far 
from having reached the precision needed to push the  
theoretical error on $\epsp/\eps$ below  $10^{-3}$ 
(assuming this is estimated in a conservative way\ldots).
In the long term,  lattice QCD seems to be much
more promising, especially in view 
of some recent theoretical developments 
in this field.\cite{Golterman}

\subsection{Isospin-breaking effects}
Violations of isospin symmetry are generated by the 
mass difference between up and down quarks and by electromagnetic 
interactions. The two types of effects are comparable 
in size and generally small in the $K$ system ($\sim 1 \%$).
In $K \to 2 \pi$ decays, however, these could be 
enhanced by a factor $1/\omega$ if an isospin-breaking (IB)
correction to $A_0$ leads to a $\Delta I > 1/2$ transition.
Thus, whereas we can safely neglect IB 
corrections proportional to $G_{27}$, and effects 
quadratic in $(m_d-m_u)$ or $\alpha$, it is important to
treat in a systematic way terms of $O[G_8 (m_d-m_u) p^n]$
and $O(G_8 e^2 p^n)$.

Once IB effects are included, the amplitude decomposition
(\ref{Adec1}) is no longer valid. In this case a useful 
parametrization is provided by\cite{Cirigliano1}
\beqa
A(K^0 \to \pi^+\pi^-) &=& A_{1/2} e^{i (\delta_0+\gamma_0)} +  
  {1\over \sqrt{2}} (A_{3/2} +A_{5/2}) e^{i (\delta_2+\gamma_2)}~, \no \\
A(K^0 \to \pi^0\pi^0) &=& A_{1/2} e^{i (\delta_0+\gamma_0) } -  
  \sqrt{2} (A_{3/2} + A_{5/2}) e^{i ( \delta_2 +\gamma_2)}~, \no \\
A(K^+ \to \pi^+\pi^0) &=& \left[  { 3 \over 2 }  A_{3/2}  - A_{5/2} \right]
  e^{i (\delta_2+\gamma^\prime_2) }~,
\label{Adec2}
\eeqa
where again the $A_i$ are real in the 
absence of CP violation.\footnote{~In 
the CP-conserving case (\protect\ref{Adec2}) 
provides the most general parametrization of three complex numbers.}

As long as $O[G_8 (m_d-m_u) p^n]$ effects are concerned, it is 
easy to show that both $\gamma_i$ and $A_{5/2}$ are 
zero to all orders. Thus the main effect induced by $m_d\not=m_u$ 
is only a shift in $A_{3/2}$, usually parametrized by 
the ratio 
\beq
\Omega_{IB} = \delta A_{3/2}/(\omega A_{1/2}). 
\eeq
At the lowest order, $O[G_8 (m_d-m_u) p^0]$, only 
the $\pi^0$--$\eta$ mixing in the strong Lagrangian ($\cL_S^{(2)}$) 
contributes to $\Omega_{IB}$, leading to the unambiguous result
$\Omega^{(2)}_{IB}=0.13$. Interestingly at this level 
$\Omega_{IB}$ is a universal correction factor, i.e. it  
applies independently to all the $\Delta I=1/2$ 
matrix elements of four-quark operators. 
Also at $O[G_8 (m_d-m_u) p^2]$ there is a 
contribution coming from the mixing on the external legs
that is universal and calculable unambiguously from 
the strong Lagrangian. This has recently been 
evaluated\cite{EMNP} and, summed to the lowest-order result, 
leads to $\Omega^{(4-mix)}_{IB}=0.16 \pm 0.03$.
However this is not the full story at $O[G_8 (m_d-m_u) p^2]$,
since at that order there appear also contributions 
from  $O(p^4)$  weak counterterms which are not universal 
and not known from data. At the moment these can only be 
estimated by means of model-dependent assumptions, and 
some recent analyses\cite{Gardner1,Wolfe} indicate sizeable 
effects, comparable in size to the one of the leading-order 
term. It should also be noted that a positive 
$\Omega_{IB}$ worsens the problem of the  $\Delta I=1/2$
rule, indicating that in the isospin limit the ratio
$|A_{2}/A_{0}|$ should be smaller than $\omega$. 
It is thus more likely that the remaining 
$O[G_8 (m_d-m_u) p^2]$ terms will decrease $\Omega_{IB}$, 
rather than enhance it, 
in agreement with the recent findings.\cite{Gardner1,Wolfe}
At the moment, in the absence of precise estimates,
what can be considered  a conservative 
approach toward phenomenological analyses is the 
use of two independent $\Omega_{IB}$ for 
CP-conserving and CP-violating parts of the amplitudes,
with a central value close $\Omega^{(2)}_{IB}$ and a 
$\sim 100\%$ error in both cases.

\medskip

Given the absence of $O[G_8 (m_d-m_u) p^n]$ 
contributions in  $A_{5/2}$ and $\gamma_i$, these 
terms are expected to be mainly of electromagnetic origin.
Present data show some evidence for these effects. Indeed 
fixing the phase difference $(\delta_0-\delta_2)$ from 
$\pi\pi$ scattering, setting $\gamma_i=0$ and fitting $K\to\pi\pi$ widths
leads to extract a non-vanishing 
$\Delta I=5/2$ amplitude:\cite{Gardner2}
$\Re(A_{5/2}/A_{1/2}) = -(7 \pm 2) \times 10^{-3}$. Interestingly 
this is of the correct order of magnitude, being $O( \alpha A_{1/2})$.
At this point, however, an important warning 
should be made concerning the unclear treatment 
of soft radiation in the $K\to\pi\pi$ data available 
at present. This issue is very 
important in this context,\cite{CiriglianoEM} and it is highly
desirable to have a clearer experimental 
information in this respect, together with more 
precise measurements of $K\to\pi\pi$ widths. 
Hopefully these should become available
in the very near future from KLOE.\cite{Franzini}

From the theoretical point of view, electromagnetic effects 
have a rather trivial structure at  $O(G_8 e^2 p^0)$, 
where there are no contributions to  $A_{5/2}$ and $\gamma_i$.
At this level there are two contributions to $\Omega_{IB}$,
one from the $\pi^+ - \pi^0$ mass difference, the other 
from the lowest-order realization of a $|\Delta S|=1$ operator transforming 
as $(8_L,8_R)$. Only the former is calculable 
unambiguously, but the cancellation of quadratic 
divergences in the photon loops is a strong indication 
that the two contributions tend to cancel each other, 
leading to an overall small effect. 
This expectation is supported by a detailed analysis of
Cirigliano {\em et al.}\cite{Cirigliano2} 
and seems to indicate that $\Omega_{IB}$ is largely 
dominated by the $m_d \not= m _u$ effects discussed before. 

The interesting aspects of electromagnetic effects 
appear at $O(G_8 e^2 p^2)$, with a non-vanishing 
$\Delta I=5/2$ amplitude and also with the bremsstrahlung 
of the leading $O(G_8 e^0 p^2)$ terms.\cite{CiriglianoEM} 
As already mentioned,
the structure of the local $O(G_8 e^2 p^2)$ operators  
and their anomalous dimensions has recently been analysed,\cite{noi} 
but a precise evaluation of their couplings is
not available yet. Despite this uncertainty some interesting 
conclusion can still be drawn. For instance 
Cirigliano {\em et al.}\cite{Cirigliano1} 
have been able to show that also at $O(G_8 e^2 p^2)$ 
electromagnetic contributions to 
$\Omega_{IB}$ are rather small (at the per cent level).
On the other hand more work is needed 
to understand the size of the $\Delta I=5/2$ amplitude,
and new precision data on $K \to \pi\pi$ widths
could be of great help in this direction.

\section{Rare $K\to \ell^+\ell^-(\pi)$ decays}
The rare processes $K\to\pi \ell^+\ell^-$ and $K_L \to \ell^+\ell^-$
are particularly interesting since  
they offer,  at the same time,
a new laboratory for understanding chiral dynamics 
and a rather clean window on the short-distance mechanism 
of flavour-changing neutral currents.
     
\subsection{$K\to \pi \ell^+\ell^-$}
The single-photon exchange amplitude ($K\to\pi \gamma^* \to 
\pi \ell^+\ell^-$) is largely dominant in these channels when
allowed by CP invariance. 
This occurs in the charged modes, 
experimentally observed for both $\ell=e$ and $\ell=\mu$,
and in the decays of the $K_S$.   
This long-distance amplitude can be described in a model-independent 
way in terms of two form factors, $W_+(z)$ and $W_S(z)$,
defined by\cite{DEIP}
\beqa
& i \dis\int d^4x e^{iqx} \langle \pi(p)|T \left\{J^\mu_{\rm elm}(x)
{\cal L}_{\Delta S=1}(0) \right\} |
K_i (k)\rangle =& \nonumber \\
&\dis\frac{W_i(z)}{(4\pi)^2}\left[z(k+p)^\mu -(1-r_\pi^2)q^\mu
\right]~, & \label{eq:tff}
\eeqa
where $q=k-p$, $z=q^2/M_K^2$ and $r_\pi = M_\pi /M_K$.
The two form factors are non-singular at $z=0$ and,
because of gauge invariance, vanish to $O(p^2)$.\cite{EPR}
Beyond lowest order one can identify 
two separate contributions to the $W_i(z)$:
a non-local term, $W_i^{\pi\pi}(z)$, due to the 
$K\to 3\pi\to \pi\gamma^*$ scattering, 
and a local term, $W_i^{\rm pol}(z)$,
that encodes contributions of the LECs.
At $O(p^4)$ the local term is simply a constant,
whereas at $O(p^6)$ also a term linear in $z$ arises.
Already at  $O(p^4)$
chiral symmetry alone does not help to 
relate $W_S$ and $W_+$, 
or $K_S$ and $K^+$ decays.\cite{EPR}

Recent results\cite{E865} on $K^+ \to \pi^+ e^+ e^-$ and
$K^+ \to \pi^+ \mu^+ \mu^-$ by E865
indicate very clearly that, owing to 
a large linear slope, the $O(p^4)$ 
expression of $W_+(z)$ is not sufficient
to describe experimental data.
This should not be considered as a failure of CHPT, 
rather as an indication that large $O(p^6)$ contributions
are present in this channel. Indeed the $O(p^6)$ expression of  
$W_+(z)$ fit the data very well; 
this is not only due to the presence of a new free parameter, 
but also  to the presence of the non-local term.
The evidence of the latter provides a real significant test 
of the chiral approach. 
It should also be stressed that the appearance of a 
large $O(p^6)$ correction in this channel can be 
qualitatively understood in terms of vector-meson exchange.
A quantitative test of this hypothesis could be performed 
with the observation of the $K_S\to \pi^0 e^+ e^-$
decay, whose branching ratio is expected in the 
$10^{-9}$--$10^{-8}$ range.\footnote{~An experimental upper limit of 
$1.6\times 10^{-7}$ has recently been obtained by NA48.\protect\cite{NA48}}

In $K_L \to \pi^0 e^+ e^-$  the long-distance part of the
single-photon exchange amplitude is forbidden by CP invariance, but it
contributes to the processes via $K_L$--$K_S$ mixing, leading to
\beq
B(K_L \to \pi^0 e^+ e^-)_{\rm CPV-ind}~=~ 3\times 10^{-3}~ B(K_S \to \pi^0 e^+
e^-)~.
\eeq
On the other hand, there is also a sizeable 
direct-CP-violating contribution to this channel
that is dominated by short-distance dynamics.\cite{GW,BLMM}
Within the SM, this theoretically clean part of the amplitude 
leads to\cite{BLMM}
\beq
B(K_L\to\pi^0 e^+e^-)^{\rm SM}_{\rm CPV-dir}~=~(2.5 \pm 0.2) \times 10^{-12}~\left[ \frac{\Im
(V_{ts}^*V_{td}) }{  10^{-4} } \right]^2,
\eeq
where $V_{ij}$ denote the elements of the 
Cabibbo--Kobayashi--Maskawa matrix. 
The two CP-violating components of the $K_L\to\pi^0 e^+e^-$ amplitude
will in general interfere with a relative phase that is 
known (the phase of $\eps$). 
Thus if $B(K_S \to \pi^0 e^+ e^-)$ will be measured, 
it will be possible to determine the interference between direct
and indirect CP-violating components of $B(K_L\to\pi^0 e^+e^-)$ 
up to a sign ambiguity.
Given the present uncertainty on $B(K_S \to
\pi^0 e^+ e^-)$  we can only, for the moment,  set a rough upper limit
of ${\rm few}\times 10^{-11}$ on the sum of all 
CP-violating contributions to this mode,\cite{DEIP} 
which is almost one order of magnitude below  
the recent experimental upper bound by KTeV.\cite{KTeV}

An additional contribution to $K_L \to \pi^0 \ell^+ \ell^-$ 
decays is generated by the CP-conserving amplitude
$K_L \to \pi^0 \gamma \gamma \to \pi^0 \ell^+ \ell^-$.\cite{Sehgal} 
Here again chiral dynamics, together with data, helps us to put 
bounds on this long-distance effect. The recent results\cite{NA48,KTeV_KLpgg} 
on $K_L \to \pi^0 \gamma\gamma$ at small $M_{\gamma\gamma}$ 
indicate that this contribution is small in the electron channel 
($\lsim~2\times 10^{-12}$). In addition the CP-conserving 
amplitude does not interfere with the CP-violating one in the 
total width and, in principle, it could be experimentally constrained 
by means of a Dalitz plot analysis. In view of these 
arguments, the CP-conserving contamination
should not represent a serious problem for  
the extraction of the interesting direct-CP-violating 
component of $B(K_L\to\pi^0 e^+e^-)$.

\subsection{$K_L \to l^+ l^-$}
The dominant contribution to these transitions, 
for both $\ell=e$ and $\ell=\mu$,  
is generated by the intermediate two-photon state. 
This leads to an absorptive amplitude 
computed unambiguously by means of $\Gamma(K_L \to \gamma\gamma)$
and a dispersive one that is more difficult to estimate,
depending on {\em a priori} unknown LECs.

In $K_L\to e^+ e^-$ the dispersive integral of 
the $K_L \to \gamma \gamma \to l^+ l^-$ loop
is dominated by a large logarithm ($\log(m_K^2/m^2_e)$),
and the relative contribution of the local term  
is small. This implies that in this case also the dispersive amplitude  
can be estimated with a relatively good accuracy 
in terms of $\Gamma(K_L\to \gamma \gamma)$,
yielding  the prediction\cite{VP} 
$B(K_L \to e^+ e^-) \sim 9 \times 10^{-12}$~, 
which has been confirmed by E871.\cite{E871}

More interesting from the short-distance point of view is the case of $K_L
\to \mu^+\mu^-$. Here the absorptive two-photon long-distance amplitude 
is not enhanced by large logs and is comparable in size
with the short-distance contribution of $Z$-penguin and $W$-box diagrams.
The latter is particularly interesting, since it is sensitive to $\Re V_{td}$
and calculable with high accuracy.\cite{BB} 
On the other hand, the dispersive part of the two-photon
contribution is more difficult to be estimated in 
this case, as it is more sensitive to the local counterterm.  

The counterterm appearing in the two-photon amplitude of 
$K_L \to \mu^+\mu^-$ is related to the behaviour 
of the $K_L \to \gamma^* \gamma^*$ form 
factor and it can be constrained by means of 
theoretical and experimental information on the 
latter.\cite{VP,BMS}$^{\rm -}$\cite{deRafmm}
To this purpose it should be stressed how important it is 
to have precise 
experimental data on  $K_L \to \gamma \ell^+\ell^-$ and
$K_L \to e^+e^- \mu^+\mu^-$ decays; these are 
also sensitive to the $K_L \to \gamma^* \gamma^*$ 
form factor,\cite{DIP,Goity} although only in a 
limited kinematical region. For instance the recent 
KTeV data\cite{Wah} suggest an inconsistency of 
the so-called BMS ansatz\cite{BMS} in fitting 
the $K_L \to \gamma e^+ e^-$ and  
$K_L \to \gamma \mu^+ \mu^-$ modes at the same time. 
Using the very precise experimental determination of 
$B(K_L \to \mu^+\mu^-)$ reported by E871,\cite{KLmm} 
together with the analysis of the two-photon dispersive 
contribution discussed in Ref.~38,  
we can obtain stringent constraints on possible extensions of the SM
and also an interesting lower bound on 
$\Re V_{td}$.\cite{KLmm} Two experimental tests that 
could strengthen (or weaken) the validity of 
this information could be obtained by means of the 
quadratic slope in the $K_L \to \gamma \ell^+\ell^-$ 
form factor (i.e. the dependence on $M_{\ell\ell}^4$)
and, especially, by means of the 
mixed slope in $K_L \to e^+e^- \mu^+\mu^-$
(i.e. the dependence on $M_{ee}^2\times M_{\mu\mu}^2$).\cite{DIP}

\section{Concluding remarks}
In the last few years there has been  major 
progress in the study of kaon decays from the 
experimental point of view. Just to mention a few examples, 
I wish to recall the observation of direct CP violation in 
$K\to 2 \pi$ decays, the evidence of the rare 
modes $K^+\to\pi^+\nu\bar{\nu}$ and $K_L \to e^+ e^-$,
the high-precision measurements of 
$K^+ \to \pi^+\pi^-  e \nu_e$ and $K^+ \to \pi^+  e^+ e^-$
form factors. Most of these results are of the utmost importance and 
have triggered a large amount of theoretical work, only
a small fraction of which has been mentioned in this talk.

On the other hand, I would like to emphasize that this field 
is far from being exhausted. There are still fundamental 
problems that require a detailed answer, such as the nature 
of the underlying mechanism of CP violation or the r\^ole 
of resonances in non-leptonic weak transitions.
These questions could be addressed by means of new experimental 
studies on $K$ decays, which have just entered
a new exciting era of precision measurements.

\section*{Acknowledgements}
I am grateful to G. Buchalla,  G. Colangelo, G. Martinelli 
and T. Pich for useful discussions. 
I am also in debt with the co-authors of Ref.~9 for 
having let me join them 
after their collaboration was already started. 
Finally I wish to thank the organizers of Chiral Dynamics 2000
for the invitation to this fruitful workshop.

\end{document}